\documentstyle[aps,amssymb,twocolumn]{revtex}

\begin{document}
\author{S.B. Rutkevich}
\address{{\it Institute of Physics of Solids}\\
{\it and Semiconductors, P.Brovki 17, Minsk, Belarus, }\\
{\it e-mail: lttt@ifttp.bas-net.by}}
\title{Decay of the metastable phase in $d=1$ and $d=2$ Ising models}

\begin{abstract}
We calculate perturbatively the tunneling decay rate $\Gamma $ of the
metastable phase in the quantum $d=1$ Ising model in a skew magnetic field
near the coexistence line $0<h_{x}<1,\ h_{z}\rightarrow -0$ at $T=0$. It is
shown that $\Gamma $ oscillates in the magnetic field $\ h_{z}$ due to
discreteness of the excitation energy spectrum. After mapping of the
obtained results onto the extreme anisotropic $d=2$ Ising model at $T<T_{c}$%
, we verify in the latter model the droplet theory predictions for the free
energy analytically continued to the metastable phase. We find also evidence
for the discrete-lattice corrections in this metastable phase free
energy.\bigskip

PACS\ numbers: 05.50, 64.60.My, 64.60.Qb, 82.60.Nh
\end{abstract}

\maketitle
\date{}


It is widely accepted (for review, see \cite{RG}) that for all $T<T_{c}$,
the free energy $F(H)$ of the two-dimensional Ising model analytically
continued from the positive real axis $H>0$ into the complex $H$-plane, has
a branch cut singularity at the origin. Near the cut drawn along the
negative real axis $H<0$, the imaginary part of the free energy is believed
to have the following form \cite{Ha} 
\begin{equation}
\text{Im }F(\text{e}^{\pm i\pi }\mid H\mid )=\mp B\mid H\mid \exp \left(
-A/\left| H\right| \right)  \label{FF}
\end{equation}
for small $\left| H\right| $. This expression extrapolates to the Ising
model the results obtained by Langer \cite{La1} and N.J. G\"{u}nther {\it et
al }\cite{G} in the droplet field theory analysis of the coarse-grained
Ginzburg-Landau model. In the droplet theory, the free energy continued to
the cut $H<0$ is interpreted as the free energy of the metastable state $%
F_{ms}(H)\equiv F($e$^{-i\pi }\mid H\mid )$. Langer conjectured \cite{La1},
that {\it Im} $F_{ms}(H)$ may be identified (up to a dynamical factor) with
the metastable phase decay rate provided by the thermally activated
nucleation.

The phenomenological droplet theory prediction for the amplitude $A$ in (\ref
{FF}) is \cite{RG,Ha,GRN} 
\begin{equation}
A=\frac{\beta \hat{\Sigma}^{2}}{8M}  \label{fd}
\end{equation}
where $M$ is the spontaneous magnetization, and $\hat{\Sigma}^{2}\ $denotes
the square of surface free-energy of the equilibrium shape droplet divided
by its area. Both $\hat{\Sigma}^{2}$ and $M$ relate to the equilibrium
zero-field state, and are known exactly. The quantity $\hat{\Sigma}^{2}$ can
be calculated by use of the Wulff's construction from the exact anisotropic
surface-tension, as it was shown by Zia and Avron \cite{ZA}, and $M$ was
obtained by Onsager and Yang \cite{Ya}. The linear depending on $\left|
H\right| $ prefactor in (\ref{FF}) arises in the continuum droplet field
theory \cite{G} from the contribution of the surface excitations (Goldstone
modes) of the critical droplet.

Equations (\ref{FF}), (\ref{fd}) were confirmed by C.C.A. G\"{u}nther,
Rikvold and Novotny \cite{GRN} in numerical constrained transfer matrix
calculations, and by Harris in numerical analysis of certain power series 
\cite{Ha}. However, no analytic microscopic evaluation of (\ref{FF}), (\ref
{fd}) for the $d=2$ Ising model did exist. The purpose of the present Letter
is to perform analytic verification of (\ref{FF}), (\ref{fd}) for the $d=2$
Ising model in the extreme anisotropic limit.

We start from the anisotropic Ising model on the square lattice. It is
characterized by the nearest neighbor in-row and in-column coupling
constants $J_{1}>0,\ J_{2}>0$, the magnetic field $H$ and the inverse
temperature $\beta =1/(k_{B}T)$. The extreme anisotropic limit of the model
is defined as follows \cite{Kog} 
\begin{eqnarray}
\tau &\equiv &\exp (-2\beta \ J_{2})\rightarrow 0,\qquad  \label{an} \\
J_{1} &=&\frac{\tau }{\beta \ h_{x}}\rightarrow 0,\qquad H=h_{z}\
J_{1}\rightarrow 0  \nonumber
\end{eqnarray}
with constant $h_{x}$ and $h_{z}$. The transfer matrix of the $d=2$ Ising
model in this limit can be written up to a nonsignificant numerical factor
as $\exp (-\beta \ J_{1}{\cal H})$, where ${\cal H}$ denotes the quantum
spin-1/2 Hamiltonian of the Ising chain in a skew magnetic field: 
\begin{equation}
{\cal H}=-\sum_{n=1}^{N}(\ \sigma _{n}^{z}\sigma _{n+1}^{z}+h_{x}\sigma
_{n}^{x}+h_{z}\sigma _{n}^{z})  \label{H}
\end{equation}
Here $\sigma ^{x,z\text{ }}$are the Pauli matrices, $N$ is the number of
sites in the chain, cyclic boundary conditions are supposed. The free energy 
$F$ per the lattice site of the two-dimensional Ising model is proportional
in limit (\ref{an}) to the ground state energy $E(h_{x},h_{z})$ of the
quantum Hamiltonian (\ref{H}): 
\begin{equation}
F=J_{1}\lim_{N\rightarrow \infty }\left[ \frac{E(h_{x},h_{z})}{N}\right]
\label{fr}
\end{equation}

Our strategy contains two steps. (i) We calculate perturbatively the
tunneling decay rate (that is, the imaginary part of the energy) of the
metastable state in model (\ref{H}) at zero temperature. (ii) By use of
mapping (\ref{fr}) we obtain then the imaginary part of the metastable free
energy $F_{ms}(H)$ at $T<T_{c\text{ }}$for the $d=2$ Ising model in limit (%
\ref{an}).

It should be noted, that models like (\ref{H}) have been widely used to
describe dynamical properties observed in real quasi-one-dimensional Ising
ferromagnets \cite{Tink,Smit}. So, the problem outlined in step (i) is by
itself of considerable physical interest.

In the small field limit $h_{x}\ll 1,$ $h_{z}\ll 1$ model (\ref{H}) was
studied by Fogedby \cite{Fog}. In the free-fermion point $h_{z}=0$
Hamiltonian (\ref{H}) reduces to the form \cite{Tsuk} 
\begin{equation}
{\cal H}_{0}\equiv {\cal H}\mid _{h_{z=0}}=\int_{-\pi }^{\pi }\frac{\text{d}%
\theta }{2\pi }\ \omega (\theta )\ \psi ^{\dagger }(\theta )\ \psi (\theta
)+Const,  \label{Ht}
\end{equation}
where $\theta $ is the quasimomentum, fermionic operators $\psi ^{\dagger
}(\theta )$, $\ \psi (\theta )$ satisfy the canonical anti-commutational
relations 
\begin{eqnarray*}
\left\{ \psi (\theta )\ ,\psi (\theta ^{\prime })\right\} &=&\left\{ \psi
^{\dagger }(\theta )\ ,\psi ^{\dagger }(\theta ^{\prime })\right\} =0 \\
\left\{ \psi ^{\dagger }(\theta )\ ,\psi (\theta ^{\prime })\right\} &=&2\pi
\delta (\theta -\theta ^{\prime })
\end{eqnarray*}
and 
\[
\omega (\theta )=2\left[ (1-h_{x})^{2}+4h_{x}\sin ^{2}\frac{\theta }{2}%
\right] ^{1/2} 
\]
At zero temperature, there is a phase transition point $h_{x}=1$, which
divides ordered ($0<h_{x}<1$) and disordered ($h_{x}>1$) phases. Two
ferromagnetic ground states $\mid 0_{+}\rangle $ and $\mid 0_{-}\rangle $
coexist in the interval $0<h_{x}<1$. They are distinguished by the sign of
the spontaneous magnetization $\langle 0\pm \mid \sigma _{n}^{z}$ $\mid
0_{\pm }\rangle =\pm $ $M$, where $M=(1-h_{x}^{2})^{1/8}$.

A small negative longitudinal magnetic field $h_{z}<0$ removes the ground
state degeneration. The following fermionic representation for the
Hamiltonian (\ref{H}) is valid in the thermodynamic limit $N\rightarrow
\infty $%
\begin{equation}
{\cal H}={\cal H}_{0}+V+Const,  \label{h1}
\end{equation}
where ${\cal H}_{0}$ denotes the free-fermionic Hamiltonian (\ref{Ht}), $V$
is given by 
\begin{eqnarray}
V &=&\mid h_{z}\mid M\sum_{n\in Z}:\exp \frac{\varrho _{n}}{2}:  \label{v} \\
\frac{\varrho _{n}}{2} &=&-\sum_{j<n}\psi _{j}^{(+)}\ \psi _{j}^{(-)} 
\nonumber \\
\psi _{j}^{(+)} &=&i\int_{-\pi }^{\pi }\frac{\text{d}\theta }{2\pi }\ \frac{%
\exp (ij\theta )}{\sqrt{\omega (\theta )}}\left( \psi (\theta )+\psi
^{\dagger }(-\theta )\right)   \nonumber \\
\psi _{j}^{(-)} &=&i\int_{-\pi }^{\pi }\frac{\text{d}\theta }{2\pi }\ \exp
(ij\theta )\sqrt{\omega (\theta )}\left( -\psi (\theta )+\psi ^{\dagger
}(-\theta )\right)   \nonumber
\end{eqnarray}
and $\psi (\theta )\mid 0_{+}\rangle =0$ for all $\theta $. We have used the
conventional notation $:...:$ for the normal ordering with respect to the
fermionic operators $\psi (\theta ),\ \psi ^{\dagger }(\theta )$.
Representation (\ref{h1}), (\ref{v}) can be obtained from (\ref{H}) by
applying the Jordan-Wigner \cite{SML} and duality \cite{Kog}
transformations. In performing the normal ordering of fermionic operators in
equation (\ref{v}) we followed to Jimbo {\it et al }\cite{JM}.

The nonlinear interaction term $V$ in (\ref{h1}) prevents the exact
integrability of the model. So, the natural way to study model (\ref{h1})
for small $\mid h_{z}\mid \neq 0$ is to use a certain perturbation
expansion. It is clear, however, that the straightforward perturbation
theory with the zero-order Hamiltonian ${\cal H}_{0}$ and perturbation $V$
is useless in the considered problem. This is due to the fact, that the term 
$V$ contains the long-range interaction $V_{0}$ between fermions, which is
given by 
\begin{equation}
V_{0}\equiv V\mid _{\omega (\theta )\rightarrow 1}=\mid h_{z}\mid
M\sum_{n\in Z}:\exp (-2\sum_{j<\ n}b_{j}^{\dagger }b_{j}):  \label{v0}
\end{equation}
where 
\begin{eqnarray*}
b_{j} &=&\int_{-\pi }^{\pi }\frac{\text{d}\theta }{2\pi }\psi (\theta )\exp
\left( ij\theta \right) \\
b_{j}^{\dagger } &=&\int_{-\pi }^{\pi }\frac{\text{d}\theta }{2\pi }\psi
\dagger (\theta )\exp \left( -ij\theta \right)
\end{eqnarray*}
Operator $V_{0}$ is diagonal in the coordinate representation: 
\begin{eqnarray}
V_{0}\ b_{j_{2n}}^{\dagger }b_{j_{2n-1}}^{\dagger }...b_{j_{1}}^{\dagger }
&\mid &0_{+}\rangle =  \nonumber \\
-h_{z}M[N-2\sum_{l=1}^{n}(j_{2l}-j_{2l-1})]\ b_{j_{2n}}^{\dagger
}b_{j_{2n-1}}^{\dagger }...b_{j_{1}}^{\dagger } &\mid &0_{+}\rangle
\label{vc}
\end{eqnarray}
where $j_{l}<j_{l+1}$. Since interaction (\ref{vc}) depends linearly on the
distance between fermions, it changes the structure of the energy spectrum
of model (\ref{h1}) at arbitrary small longitudinal magnetic field $%
h_{z}\neq 0$. So, to describe decay of the metastable vacuum, one should
include the long range interaction $V_{0}$ into the zero-order Hamiltonian.
This phenomenon is well known in the Stark effect \cite{LL}. To describe
ionization of an atom by the uniform electric field ${\bf E}$, one needs to
consider the corresponding electrostatic energy $e{\bf Er}$ in a
non-perturbative way.

Accordingly, we subdivide the Hamiltonian (\ref{h1}) into the zero-order and
interaction parts, as follows 
\begin{equation}
{\cal H}=\tilde{{\cal {H}}}_{0}+\tilde{V}  \label{h2}
\end{equation}
where 
\begin{eqnarray}
\tilde{{\cal {H}}}_{0} &\equiv &{\cal H}_{0}+V_{0}+Const,  \label{0h} \\
\qquad \tilde{V} &\equiv &V-V_{0}  \label{vt}
\end{eqnarray}
The numerical constant in (\ref{0h}) in chosen to provide$\tilde{\text{ }%
{\cal {H}}}_{0}\mid 0_{+}\rangle =0$. Since the new zero-order Hamiltonian $%
\tilde{{\cal {H}}}_{0}$ conserves the number of fermions, its eigenstates
can be classified by the fermion number. It is clear from (\ref{vc}), that
fermions created by operators $b_{n}$ are just the domain walls dividing the
chain into oppositely magnetized domains.

One can easily verify in the small $h_{x}$-limit, that the metastable vacuum 
$\mid 0_{+}\rangle $ decays preferably into a one-domain state. We suppose,
that this is true also in the general case $0<h_{x}<1$. So, below we shall
contract the space of considered states to the two-fermion (i.e. one-domain)
sector.

Let $\mid \phi _{l}\rangle $ be the translation invariant two-fermion
eigenstate of the Hamiltonian $\tilde{{\cal {H}}}_{0}$. In the coordinate
representation, the zero-order eigenvalue problem $\tilde{{\cal {H}}}%
_{0}\mid \phi _{l}\rangle =E_{l}\mid \phi \rangle $ takes the form 
\[
\sum_{n^{\prime }\in Z}K_{nn^{\prime }}\ \phi _{l}(n^{\prime })-M\mid n\
h_{z}\mid \phi _{l}(n)=\frac{E_{l}}{2}\phi _{l}(n) 
\]
where 
\begin{eqnarray*}
K_{nn^{\prime }} &=&\int_{-\pi }^{\pi }\frac{\text{d}\theta }{2\pi }\ \omega
(\theta )\exp [i(n-n^{\prime })\theta ] \\
\phi _{l}(n) &=&\langle 0_{+}\mid b_{0}\ b_{n}\mid \phi _{l}\rangle ,\qquad
\phi _{l}(-n)=-\phi _{l}(n)
\end{eqnarray*}
If the energy $E_{l}$ is small enough, $E_{l}<\varepsilon $, where $%
\varepsilon \ll \omega (0)$, the wavefunction $\phi _{l}(n)$ is mainly
concentrated far from the origin in the classically available region $\mid
n\mid >\omega (0)/(\mid h_{z}\mid M)$. Therefore, we can apply the `strong
coupling approximation' \cite{Zim} to represent the wavefunction in the form 
\begin{equation}
\phi _{l}(n)\cong \varphi _{l}(n)-\varphi _{l}(-n)  \label{str}
\end{equation}
where the function $\varphi _{l}(n)$ solves the equation 
\[
\sum_{n^{\prime }\in Z}K_{nn^{\prime }}\ \varphi _{l}(n^{\prime })-\mid
h_{z}\mid M\ n\ \varphi _{l}(n)=\frac{E_{l}}{2}\varphi _{l}(n) 
\]
\newline
\newline
After the Fourier transform, we obtain 
\[
\varphi _{l}(n)=\int_{-\pi }^{\pi }\frac{\text{d}\theta }{2\pi }\varphi
_{l}(\theta )\exp \left( in\theta \right) 
\]
where 
\begin{eqnarray}
\varphi _{l}(\theta ) &=&C\ \exp \left\{ -\frac{i}{2\mid h_{z}\mid M}\left[
f(\theta )-E_{l}\ \theta \right] \right\}  \label{fii} \\
C &=&(2\mid h_{z}\mid MN)^{-1/2}  \nonumber \\
f(\theta ) &=&2\int_{0}^{\theta }\text{d}\alpha \ \omega (\alpha )  \nonumber
\end{eqnarray}
The $2\pi $-periodicity condition for the function $\varphi _{l}(\theta )$
determines the energy levels $E_{l}$:

\begin{equation}
E_{l}=\frac{f(\pi )}{\pi }-2\mid h_{z}\mid \ M\ l  \label{El}
\end{equation}
The normalization constant $C$ in (\ref{fii}) is chosen to yield 
\[
\langle \phi _{l}\mid \phi _{l^{\prime }}\rangle =\frac{\delta _{ll^{\prime
}}}{\Delta E} 
\]
where $\Delta E=2\mid h_{z}\mid M$ is the interlevel distance.

To determine the decay rate $\Gamma $ of the metastable vacuum we use the
following nonregorous formal procedure. In the second order correction $E_{ms%
\text{ }}^{(2)}$ to the metastable vacuum energy we shift the excitation
energy levels downwards into the complex $E$-plane: 
\begin{equation}
E_{ms\text{ }}^{(2)}=-\Delta E\sum_{l}\frac{\mid \langle 0_{+}\mid \tilde{V}%
\mid \phi _{l}\rangle \mid ^{2}}{E_{l}-i\gamma }  \label{E2}
\end{equation}
The width $\gamma $ describes phenomenologically the decay rate of
one-domain states $\mid \phi _{l}\rangle $. Decay of these states can be
caused both by term (\ref{vt}) and by other interactions not included into
the Hamiltonian (\ref{H}). As the result, the metastable vacuum energy gains
the imaginary part

\begin{equation}
\text{Im }E_{ms}\cong -\pi \ g(h_{z})\mid \langle 0_{+}\mid \tilde{V}\mid
\phi _{l}\rangle \mid _{\text{ }E_{l}=0}^{2}  \label{ee}
\end{equation}
where 
\[
g(h_{z})=\text{Im}\cot \left[ \frac{f(\pi )-i\pi \gamma }{2\ \mid h_{z}\mid M%
}\right] 
\]
In deriving (\ref{ee}) we have extracted from the sum the slowly depending
on $l$ factor $\mid \langle 0_{+}\mid \tilde{V}\mid \phi _{l}\rangle \mid _{%
\text{ }}^{2}$ in the righthandside of (\ref{E2}).

The metastable vacuum relaxation rate $\Gamma $ is determined then in the
usual way 
\begin{equation}
\Gamma =-2\text{ Im }E_{ms}  \label{img}
\end{equation}
It is evident from (\ref{ee}), (\ref{img}) that $\Gamma $ oscillates in $%
h_{z}^{-1}$ with the period $2\pi M/f(\pi )$. These oscillations become
considerable in the case of coherent tunneling $\gamma \lesssim \Delta E$.
In the limit of noncoherent tunneling $\gamma \gg \Delta E$ oscillations in $%
h_{z}^{-1}$ vanish and relations (\ref{ee}), (\ref{img}) transform to the
Fermi's Gold Rule \cite{LL,RS}: 
\begin{equation}
\Gamma =2\pi \mid \langle 0_{+}\mid \tilde{V}\mid \phi _{l}\rangle \mid _{%
\text{ }E_{l}=0}^{2}  \label{GF}
\end{equation}
In the limit $h_{z}\rightarrow -0$ the matrix element of the interaction
operator can be asymptotically written as 
\begin{eqnarray}
\langle 0_{+} &\mid &\tilde{V}\mid \phi _{l}\rangle \cong i\mid h_{z}\mid
MN\int_{-\pi }^{\pi }\frac{\text{d}\theta }{2\pi }\ \varphi _{l}(\theta )%
\frac{\text{d}\ln \left[ \omega (\theta )\right] }{\text{d}\theta }\cong 
\nonumber \\
&&\frac{1}{3\sqrt{2}}\left( \mid h_{z}\mid MN\right) ^{1/2}\exp \left\{ -%
\frac{\mid f(\theta _{0})\mid }{2\mid h_{z}\mid M}\right\}  \label{m2}
\end{eqnarray}
where $\theta _{0}=i\mid \ln h_{x}\mid $ is the imaginary zero of the
function $\omega (\theta )$, $\omega (\theta _{0})=0$. Substitution of (\ref
{m2} ) into (\ref{ee}) yields finally 
\begin{equation}
\text{ Im }E_{ms}=-\frac{\pi }{18}N\ \mid h_{z}\mid M\ g(h_{z})\exp \left\{ -%
\frac{\mid f(\theta _{0})\mid }{\mid h_{z}\mid M}\right\}  \label{GG}
\end{equation}
Perhaps, described by (\ref{img}), (\ref{GG}) oscillations in $h_{z}$ of the
metastable state decay rate could be observed (indeed, in somewhat modified
form) in real quasi-one-dimensional Ising ferromagnets at very low
temperatures.

Now let us map obtained results to the $d=2$ Ising model. Applying (\ref{fr}%
) to (\ref{GG}) we obtain 
\begin{equation}
\text{Im }F_{ms}=B\mid H\ \mid \tilde{g}(H)\exp \left( -A/\mid H\mid \right)
\label{imF}
\end{equation}
where 
\begin{eqnarray}
\tilde{g}(H) &=&\text{Im}\cot \left\{ \frac{J_{1}\left[ f(\pi )-i\pi \gamma
\right] }{2\ \mid H\mid M}\right\}  \label{g2} \\
A &=&\frac{J_{1}}{M}\mid f(-i\ln h_{x})\mid  \label{a2} \\
B &=&\frac{\pi }{18}M  \label{b2}
\end{eqnarray}
and $H\rightarrow -0.$ These expressions should be compared with the droplet
theory predictions (\ref{FF}), (\ref{fd}).

First, let us verify, that expressions (\ref{fd}) and (\ref{a2}) for the
amplitude $A$ are equivalent. To do this, we need to determine the quantity $%
\hat{\Sigma}^{2}$ in limit (\ref{an}).

The droplet equilibrium shape in the $d=2$ Ising model is described by the
equation \cite{ZA} 
\begin{equation}
a_{1}\cosh (\beta \lambda x_{1})+a_{2}\cosh (\beta \lambda x_{2})=1
\label{sh}
\end{equation}
where $x_{1},x_{2}$ denote Descartes coordinates of a point on the droplet
boundary, the scale parameter $\lambda $ determines the droplet size, and 
\[
a_{1}=\frac{\tanh (2\beta J_{2})}{\cosh (2\beta J_{1})}\ ,\;\qquad a_{2}=%
\frac{\tanh (2\beta J_{1})}{\cosh (2\beta J_{2})} 
\]
In the extreme anisotropic limit 
\[
a_{1}\cong 1-2\tau ^{2}(1+h_{x}^{-2}),\;\qquad a_{2}\cong 4\tau ^{2}/h_{x} 
\]
and (\ref{sh}) simplifies to 
\[
x_{1}=\pm \frac{J_{1}}{\lambda }\omega \left( i\beta \lambda x_{2}\right) 
\]
Integrating in $x_{2}$ this equation we obtain the area of the equilibrium
shape droplet $S(\lambda )=W/\lambda ^{2}$, where 
\[
W=\frac{2J_{1}}{\beta }\left| f(-i\ln h_{x})\right| 
\]
It follows from the Wulff's theorem \cite{ZA}, that the surface energy $%
\Sigma (\lambda )$ also can be expressed in $W$: $\Sigma (\lambda
)=2W/\lambda $. Therefore, $\hat{\Sigma}^{2}=4W$, and 
\[
A=\frac{\beta \hat{\Sigma}^{2}}{8M}=\frac{J_{1}}{M}\mid f(-i\ln h_{x})\mid 
\]
in exact agreement with (\ref{a2}).

Further, expression (\ref{imF}) differs from (\ref{FF}) by the oscillating
factor $\tilde{g}(H)$. We interpret this factor as the correction coursed by
the discrete-lattice effects. Those may be significant at low temperatures
in the presence of strong anisotropy \cite{O}. The following observation
supports such interpretation.

At low temperatures ($h_{x}\rightarrow 0$), the factor $\tilde{g}(H)$ can be
written as 
\begin{equation}
\tilde{g}(H)=\text{Im}\cot \left\{ \pi \cdot 2x_{1}(H)-\frac{i\pi \ \gamma \
J_{1}}{2\ \mid H\mid M}\right\}  \label{g3}
\end{equation}
where $2x_{1}(H)=2J_{1}/(M\left| H\right| )$ is the continuum nucleation
theory value of the critical droplet diameter in the $x_{1}$-direction.
Maximum points in (\ref{g3}) just correspond to discrete values of this
diameter.

The oscillatory factor $\tilde{g}(H)$ contains parameter $\gamma $ which
remains undetermined in the present incomplete theory. One would expect,
however, that in the critical region ($h_{x}\rightarrow 1$) parameter $%
\gamma $ is large enough, so that $\tilde{g}(H)\cong 1$, and oscillations in
(\ref{imF}) vanish. Really, in this limit spectrum (\ref{El}) becomes
continuous, and $\Gamma $ can be obtained directly from (\ref{GF}) without
referring to (\ref{E2}), (\ref{ee}).

In the critical region expression (\ref{b2}) agrees with our previous result 
\cite{Ru}.

This work is supported by the Fund of Fundamental Investigations of Republic
of Belarus.

\end{document}